\documentclass[twocolumn,floatfix,prl,a4paper]{revtex4}

\usepackage{color}
\usepackage{graphicx}
\usepackage{amsmath}
\usepackage{bm}
\newcommand{\ket}[1]{\left|#1\right>}
\newcommand{\bra}[1]{\left<#1\right|}

\begin{document}

\title{Ground state overlap knows about ultraviolet dynamics: results for the Kondo model}
\author{C. D. Pelwan}
\author{I. Snyman}
\email{izaksnyman1@gmail.com}
\affiliation{Mandelstam Institute for Theoretical Physics, School of Physics, University of the Witwatersrand, PO Box Wits, Johannesburg, South Africa}
\date{10 December 2018}
\begin{abstract} 
We consider a quantum quench from the strongly correlated ground state of the Kondo model, to a Fermi sea. We calculate the 
overlap between the ground states before and after the
quench, as well as the Loschmidt echo, i.e. the transition amplitude between the initial state and the evolved state at a time $t$ after the quench. 
The overlap is known to determine the dynamics of the echo at large times. We show in addition that the overlap depends algebraically on the emergent Kondo length
with a power law exponent that is the difference of long and short time contributions that appear in the echo. Our result suggests that there may in general be more information contained
in the overlap than previously recognized.
\end{abstract}
\maketitle

When a many-body hamiltonian is suddenly quenched,
the system's full response cannot always be calculated perturbatively. A quantity
that captures this singularity is the
overlap between the ground states before and after the quench. 
Due to its fundamental role in non-equilibrium dynamics \cite{Munder} and quantum phase transitions \cite{Gu},
ground state to ground state overlaps have been calculated in settings
ranging from quantum impurity models \cite{Weichselbaum, Lukyanov1, Lukyanov2} through single-particle and many-body localized systems \cite{Khemani, Deng, Vasseur1}, to 
spin chains \cite{Zanardi, Cozzini} and lattices \cite{Zhou}, Luttinger liquids \cite{Gogolin}, and particles with fractional exclusion statistics \cite{Ares}, using techniques that include conformal field theory, the numerical renormalization
group, matrix product states and tensor networks.

Since the overlap compares ground states, it is not surprising that it
provides information about infrared (IR) physics, such as long time dynamics.
For instance, if the interaction between a Fermi liquid and a local impurity is suddenly quenched, 
the transition amplitude between the initial ground state, and the time evolved state after the quench -- the Loschmidt echo -- decays
algebraically at long times, with an exponent equal to that governing the system size dependence of the overlap \cite{Yuval, Othaka, Hopfield, Munder, Vasseur2, Kennes}.
However, unlike truly infrared probes, the ground state to ground state overlap is determined by the whole Fermi sea, not just the Fermi surface \cite{Lukyanov1}.
It must therefore also contain information about properties beyond the infrared.
We believe that this aspect of the overlap has been neglected because of a too narrow focus
on the overlap's dependence on the system size -- an infrared scale. To remedy this oversight,
we study a model with an emergent scale, and consider the overlap's dependence on this scale.
We are not the first to do so \cite{Lukyanov1, Lukyanov2}, but the new ingredient in our study is the link we
establish to non-equilibrium dynamics beyond the infrared. 

We consider a one-dimensional Fermi sea,
coupled to a spin-1/2 magnetic moment via a local exchange interaction
$J_\parallel S_z(0) s_z+ J_\perp \left[S_x(0)s_x+S_y(0)s_y\right]$.
Here $\vec{S}(0)$ is the electron spin density at $x=0$, and $\vec{s}$ is the impurity spin.
The interaction famously generates an emergent scale, the Kondo temperature $T_k$.
The impurity spin and the Fermi sea form a spin singlet,
in which the impurity is screened by the electron gas at distances larger than the Kondo length $\xi=\hbar v_F/2T_k$, where $v_F$ is the Fermi velocity. 
Below we set $\hbar=v_F=1$.
We consider quenches $(J_\parallel,J_\perp)\to(J_\parallel',0)$. The vanishing transverse coupling $J_\perp'=0$ in the final hamiltonian
produces a diverging Kondo length $\xi'\to\infty$. This is associated with imperfect screening of the impurity spin, and
the initial and final hamiltonians therefore have distinct infrared properties. 
Quenches in the Kondo and related models have been studied numerous times before, both theoretically
\cite{Vasseur2, Kennes, Nordlander, Anders1, Lobaskin, Anders2, Heyl, Medvedyeva, Lechtenberg, Ghosh, Andrade} and experimentally \cite{Latta}, 
but the connection between the ground state to ground state overlap and dynamics beyond the infrared has not.

Using a combination of analytical and numerical techniques, we
calculated the Loschmidt echo $P(t)=\left<J_\parallel,J_\perp\right|e^{-i H_{J_\parallel',0} t}\left|J_\parallel,J_\perp\right>$,
where $\ket{J_\parallel,J_\perp}$ is the ground state before and $H_{J_\parallel',0}$ the Hamiltonian after the quench.
We found distinct power laws in the regimes $t\ll\xi$ and $t\gg \xi$, i.e.
\begin{equation}
\left|P(t)\right|\propto\left\{\begin{array} {ll} \left(\frac{a}{t}\right)^{\chi/2}&\mbox{ if }a\ll t\ll \xi,\\
\left(\frac{a}{\xi}\right)^{\chi/2}\left(\frac{\xi}{t}\right)^{\beta/2}&\mbox{ if }\xi\ll t\ll L,\end{array}\right.\label{mainres2}
\end{equation}
where $L$ is the system size, and $a$ is a short distance cutoff of the order of the Fermi wavelength. 
The exponent $\beta$ that governs the IR dynamics ($t\gg \xi$) equals the exponent that appears in
the Anderson orthogonality theorem, in agreement with previous studies \cite{Lukyanov1, Munder}, and depends on $J_\parallel'$ but not
on the parameters of the pre-quench hamiltonain, provided $J_\perp$ is finite. In contrast, the UV exponent depends
on both $J_\parallel$ and $J_\parallel'$. Next we calculated the ground state to ground state overlap
$Q\equiv \left|\left<J_\parallel,J_\perp\right|\left. J_\parallel',0\right>\right|^2$
, and found
\begin{equation}
Q=C(J_\parallel,J_\parallel')\left(\frac{a}{\xi}\right)^{\chi/2}\left(\frac{\xi}{L}\right)^{\beta/2}.\label{mainres1}
\end{equation}
As is known already  \cite{Weichselbaum, Munder}, the $L$ dependence of the overlap is governed by the same exponent
that governs the IR dynamics of the echo. 
In contrast, the overlap's $\xi$ dependence at fixed $J_\parallel$ reveals a new power law, with an exponent equal to the difference between the UV and IR exponents
of the echo. This is our main result.   

Previous studies \cite{Lukyanov1,Lukyanov2} considered the quench $(J_\parallel,J_\perp)\to(J_\parallel,J_\perp')$, i.e. $J_\parallel$ 
was held fixed during the quench, and in general the Kondo length was
quenched between finite values $\xi$ and $\xi'$.
 A very nontrivial analytical result was obtained for the case of a finite post-quench Kondo length. However,
in the limit
$\xi'\gg\xi$, i.e. quenching to $J_\perp'=0$, 
this result reduces to the simple power law $\left|\left<J_\parallel,J_\perp\right|\left. J_\parallel',J_\perp'\right>\right|^2\propto (\xi/\xi')^{\beta/2}$,
reminiscent of the overlap between Fermi seas, with $\xi'$ playing the role of system size, and $\xi$ that of Fermi wavelength.
The $(\xi/t)^{\beta/2}$ factor in our result for $P(t)$ is consistent this interpretation: it coincides with the IR response of a Fermi sea
with Fermi wavelength $\xi$.
In  \cite{Lukyanov1,Lukyanov2}, no ultraviolet contribution to the power law exponent for $\xi$ was found.
Consistent with this, we find that up to finite size errors, the
ultraviolet contribution $\chi$ in Eq.\;(\ref{mainres1}) vanishes when we choose $J_\parallel=J_\parallel'$.
For the overlap to be sensitive to ultraviolet physics, it is therefore essential to have $J_\parallel\not=J_\parallel'$.
It is tempting to interpret the factor $(a/\xi)^{\chi/2}$ in Eq.\;(\ref{mainres1}) and the factor $(a/t)^{\chi/2}$ in Eq.\;(\ref{mainres2}) 
in terms of Anderson's orthogonality theorem, and conclude that the physics above the Kondo scale  
also allows an effective description in terms of (phase-shifted) Fermi seas.
In the UV effective theory, $\xi$ would play the role of an effective system size.
For sufficiently large $J_\parallel$,
we can show analytically that this picture is indeed correct. In this regime the exponent $\chi$ approximately equals the power law exponent one
would obtain by applying the Anderson orthogonality theorem to the trivial initial state obtained by setting $J_\perp=0$ in the pre-quench Hamiltonain. 
However, for smaller $J_\parallel$, the analytical argument breaks down, and we find numerically that $\chi$ start to deviate from the $J_\perp\to0$
result. In the absence of analytical results, we cannot prove conclusively that
a UV Fermi sea picture is correct, but if it is, there is a renormalization of $J_\parallel$ from its bare value. 
  
In our analysis, we exploit the exact mapping known to exist between the Kondo model and the ohmic spin-boson model \cite{Guinea, Kotliar, Costi}.
The spin-boson model describes a two-level system linearly coupled to a bath of harmonic oscillators. The hamiltonian is
\begin{equation}
H_{\alpha,\Delta}=\sum_{n=1}^\infty \omega_n b_n^\dagger b_n -\sqrt{\alpha}
\sum_{n=1}^\infty \frac{g_n}{2}\left(b_n+b_n^\dagger\right)\sigma_z+\frac{\Delta}{2} \sigma_x,\label{ham}
\end{equation}
where $\sigma_x$ and $\sigma_z$ are Pauli matrices, and $b_n$ are bosonic annihilation operators. We take $\Delta>0$ and denote
the ground state of $H_{\alpha,\Delta}$ as $\ket{\alpha,\Delta}$.
In the thermodynamic limit, the bath spectrum becomes dense, and the bath is completely characterized by the spectral function
$J(\omega)=\sum_{n=1}^\infty g_n^2\delta(\omega-\omega_n)$.
Typically, a power law spectral function is considered. We take
\begin{equation}
J(\omega)=2 \omega_0^{1-s} \omega^s e^{-a\omega},\label{plj}
\end{equation}
where $ e^{-a\omega}$ is a soft ultraviolet cutoff.
The ohmic model, which is equivalent to the Kondo model, has $s=1$, i.e. a linear spectral density at low frequency.
The parameter $\alpha$ is determined by $J_\parallel$ in the following way.
Let $\varphi$ be the difference in phase shifts between spin-up and spin-down electrons at the Fermi energy, 
when instead of the exchange interaction, our Fermi sea is subjected to a spin-dependent static potential $J_\parallel S_z(0)$.
Then $\alpha=(\varphi/\pi-1)^2$.

In the thermodynamic limit, one introduces Dirac-delta normalized bosonic field operators
$\phi_\omega=\lim_{L\to\infty}\sqrt{L/2 \pi}b_n$ and calculates the Kondo length as follows \cite{Snyman}
\begin{align}
\xi&=\lim_{\omega\to0}\bra{\alpha,\Delta}\frac{\phi_\omega+\phi_\omega^\dagger}{\sqrt{2\omega}}\sigma_z\ket{\alpha,\Delta}\label{klexp}.
\end{align}
In the fermionic language the above definition translates to  
$\xi=\lim_{x\to\infty}4x^2\left<S_z(x) s_z\right>$.
The familiar Kondo temperature, defined via the static spin-susceptibility \cite{Hanl} is related to the Kondo length as $T_K= 1/2 \xi$ \cite{Blunden-Codd}. 
For the Kondo quenches we are interested in, we must evaluate the spin-boson Loschmidt echo
\begin{equation}
P(t)=e^{itE_{\alpha',0}}\bra{\alpha,\Delta}e^{-i H_{\alpha',0} t}\ket{\alpha,\Delta},
\end{equation}
where $E_{\alpha',0}$ is the ground state energy of $H_{\alpha', 0}$. 
Here $\alpha'=(\varphi'/\pi-1)^2$ and the relationship between $\varphi'$ and $J_\parallel'$ is the same as that between $\varphi$ and $J_\parallel$.

Our analysis is based on an exact expansion of the ground state of $H_{\alpha,\Delta}$ in terms of a {\em discrete} set of coherent states.
The theory behind the expansion was first set out in Refs. \cite{Bera1, Bera2, Snyman}. Here and in the supplementary
material, we give a brief review.
According to an elementary theorem by Cahill \cite{Cahill}, an arbitrary state $\ket{\psi}$ of a set of bosonic modes can be written as a {\em discrete} sum
$\ket{\psi}=\sum_{m=1}^\infty c_m \ket{f_m}$
where $\ket{f_m}$ is a multi-mode coherent state, i.e. $b_n \ket{f_m}=f_{mn}\ket{f_m}$.
The representation is not unique, and there is always sufficient freedom to choose the coefficients $f_{mn}$ real.
The spin-boson hamiltonian is invariant with respect to the unitary transformation
$T$ that sends $b_n\to -b_n$ and $\sigma_z\to -\sigma_z$. 
We confine our attention to the so-called delocalized phase, where the ground state $\ket{\alpha,\Delta}$ transforms
under $T$ as $T\ket{\alpha,\Delta}=-\ket{\alpha,\Delta}$, so that $\left<\sigma_z\right>=0$. For the ohmic case ($s=1$), the
interval $\alpha\in(0,1)$ is within the delocalized phase. 
With the help of Cahill's theorem, we can then exactly parameterize the ground state as
\begin{equation}
\ket{\alpha,\Delta}=\sum_{m=1}^\infty \frac{c_m}{\sqrt{2}}\left(\ket{f_m}\ket{\uparrow}-\ket{-f_m}\ket{\downarrow}\right).\label{parametrization}
\end{equation}
In order to put this result to practical use, the sum in Eq.\;(\ref{parametrization}) is truncated to a finite number of $M$ terms
and the optimal parameters $c_m$ and $f_{mn}$ are found via the variational principle. 
Any desired accuracy can in principle be attained, by taking $M$ sufficiently large.
In the limit of small $\alpha$, the single term expansion $M=1$ 
becomes exact \cite{Silbey, Harris, Bera1}.
Significant  headway can be made analytically with the minimization, as we review in the supplementary material, and
as a result one is eventually left with only $M^2+M-1$ parameters whose values must be found via numerical minimization. 
However, some analytical results can be obtained without explicitly calculating the optimal values of these parameters.
An example is a result that we derive in the supplementary material, namely that at large times 
\begin{equation}
P(t)\simeq\left|\left<\alpha,\Delta \right.\left|\alpha',0\right>\right|^2\exp\frac{\alpha'}{4}\int_0^\infty d\omega \frac{J(\omega)}{\omega^2}e^{-i\omega t}\label{pt2}.
\end{equation}
Whereas $P(t)$ is regular in the thermodynamic limit, if the individual factors on right of Eq.\;(\ref{pt2})
are calculated separately, the calculations must be performed for a finite system size.
The time-dependence of Eq.\;(\ref{pt2}) is the same as
$\exp(iE_{\alpha',0}t)\bra{\rm IR}\exp(-i H_{\alpha',0}t)\ket{\rm IR}$, 
with $\ket{\rm IR}=\ket{0}\left(\ket{\uparrow}-\ket{\downarrow}\right)/\sqrt{2}$.
This is because in the delocalized phase, the coherent state parameters $f_{mn}$
tend to zero for modes $n$ whose frequencies $\omega_n$ tend to zero \cite{Blunden-Codd}.
As a result, infrared probes cannot distinguish the initial state from the bosonic vacuum.
Specializing now to the ohmic case, we find at times large compared to $\xi$, that 
\begin{equation}
P(t)=Q\left(\frac{2\pi i t}{L}\right)^{-\beta/2},\label{plt}
\end{equation}
with $\beta=\alpha'$ and
where we use the Kondo and spin-boson notations $Q\equiv \left|\bra{\alpha,\Delta}\left.\alpha', 0\right>\right|^2$
interchangeably.
Since $P(t)$ should not vanish when we take the limit $L\to\infty$, we conclude that
$Q\propto L^{-\beta/2}$. In the supplementary material we comment further on these known results, and explain how they 
relate to the Anderson orthogonality theorem. 

Now we turn to the Kondo length dependence of the overlap. 
We first present a heuristic argument by means of which the numerical results we subsequently present may be anticipated. 
Provided $\xi\gg a$, 
we expect $\xi$ to set the scale for
dynamics at times sufficiently larger than $a$.
This motivates the scaling Ansatz
\begin{equation}
P(t)=Z(\alpha,\alpha',\xi/a) F_{\alpha,\alpha'}\left(\frac{t}{\xi(\alpha,\Delta,a)}\right),\label{scaling}
\end{equation}
in which all parameter dependencies have been rendered explicit.
If this Ansatz is correct, we can extract $Z$ from the long time result (\ref{plt})
and obtain
\begin{align}
Z&=\left(\frac{2\pi \xi}{L}\right)^{-\beta/2}Q.
\label{eqz}
\end{align}
Now, consider the behavior of $P(t)$ at times short compared to $\xi$. We expect the degrees
of freedom that are relevant in this regime to be ignorant of physics at the Kondo scale. For
large $J_\parallel$, i.e. small $\alpha$, we can show analytically using the $M=1$ expansion 
that for $a\ll t \ll \xi$, 
\begin{equation}
P(t)=C\left(\frac{i t}{a}\right)^{-\chi/2},\label{spw}
\end{equation}
with $\chi=(\varphi/\pi-\varphi'/\pi)^2=(\sqrt{\alpha}-\sqrt{\alpha'})^2/2$ and $C$ independent of $\xi$. (See the
Supplementary Material for a derivation.)
This is the same behavior as we would have obtained if, before the quench, we had $J_\perp=0$,
and hence $\xi=\infty$. Our analytical argument however breaks down at larger $\alpha$. 
Let us nonetheless assume that a short time power law of the form (\ref{spw}) holds for arbitrary $\alpha$, 
with a $\xi$-independent $C$. 
Universal scaling according to the Ansatz (\ref{scaling}) implies that  
$P(t)/Z$
only depends on $t$ and $\xi$ in the combination $t/\xi$. The $\xi$-independence of $C$ then implies that
$Z\propto \xi^{-\chi/2}$.
Using our analytic expression (\ref{eqz}) for $Z$, we arrive at the result we want, namely
$Q \propto \xi^{(\beta-\chi)/2}$. 
Below Eq.\;(\ref{plt}) we concluded that the overlap $Q$ scales with system size as $L^{-\beta/2}$.
Since the only other length scale that the $Q$ can depend on is $a$, 
we reproduce Eq.\;(\ref{mainres1}), our main result.

\begin{figure}[tbh]
\begin{center}
\includegraphics[width=.9\columnwidth]{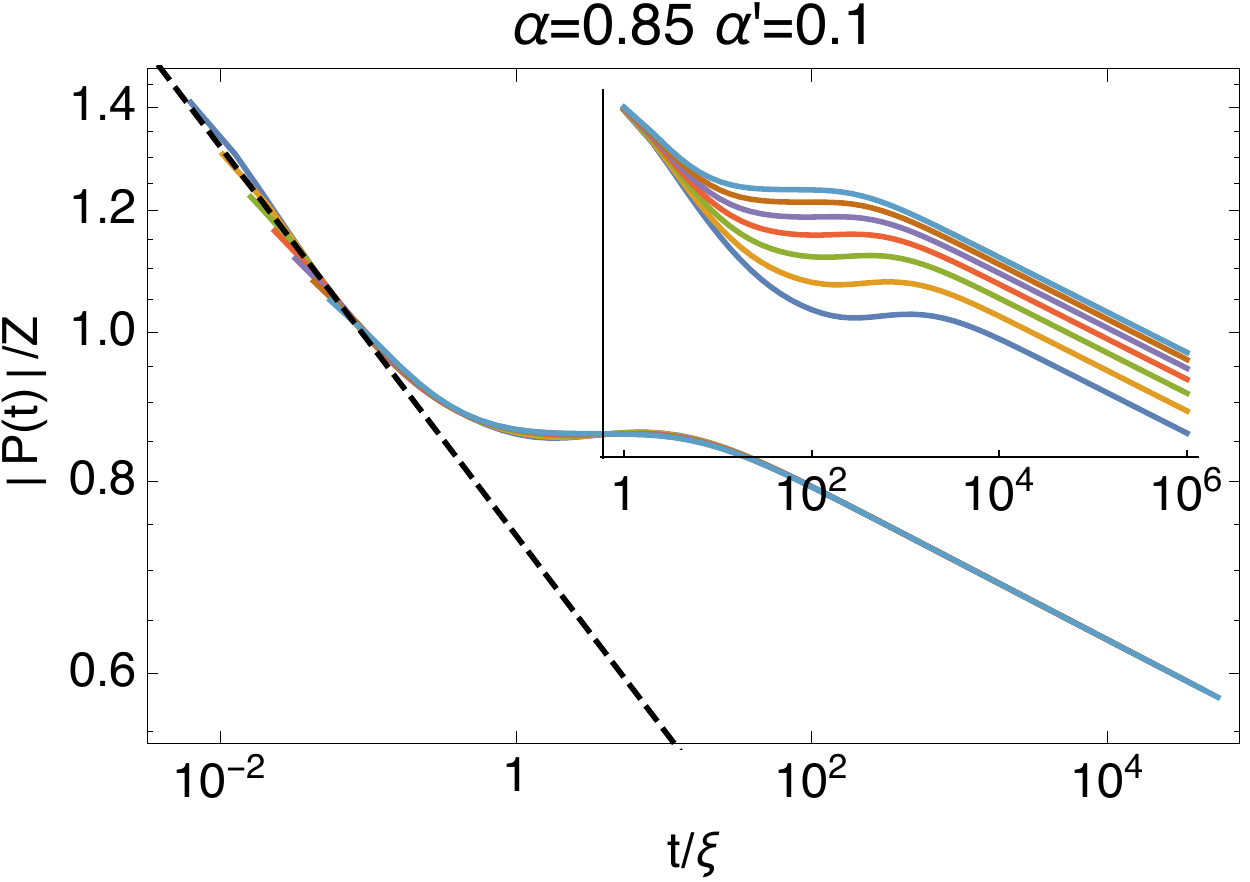}
\caption{The computed Loschmidt echo $P(t)$ for $\alpha=0.85$ and $\alpha'=0.1$. The different curves correspond different $\Delta/a$ values
$\in\{0.156,\,0.18,\,0.205,\,0.23,\,0.255,\,0.28,\,0.305\}$. The main panel shows the amplitude of the scaled function
$P(t)/Z$ v. $t/\xi$ with $Z$ and $\xi$ calculated from Eqs.\,(\ref{klexp}) and (\ref{eqz}), not fitted. The inset shows the unscaled data v. time in
units of $a$. The sloped dashed line represents the short time power law (\ref{spw}) with $C$ and $\chi$
obtained by fitting to the amplitude data for $t<10^{-1}\xi$. \label{f2}}
\end{center}
\end{figure}

In the remainder of this work, we present our numerical study, based on the coherent state expansion.
We note that a powerful extension of the technique has been developed to deal with general time dependent problems \cite{Gheeraert1,Gheeraert2}.
However, due to the simple form of the post-quench hamiltonian,
no sophisticated techniques are needed here, once the initial state is expressed in terms of the bosons.
We reuse the numerical ground state data
of \cite{Snyman}. The data were 
obtained by simulated annealing for an expansion truncated to $M=7$ terms at most. In \cite{Snyman}, 
multiple checks were done to verify that numerical convergence is satisfactory. In the supplementary material
we show how the curves for $P(t)$ calculated at increasing $M$
rapidly converge.
We estimate that the error in $P(t)$ is at most on the order of $1\%$ at large times, and less at shorter times.
For given $\alpha$, we have data for $\xi/a$ ranging over between $1$ and $2$ decades, with the shortest $\xi$ an order of magnitude larger than 
the short distance cutoff $a$.
For each of the $50$ combination of $(\alpha,\Delta)$ at which data was taken, we calculated the Loschmidt echo and the scaling factor $Z$ 
for quenches to various $\alpha'\in[0,1.5]$. Details are provided in the supplementary material.
In Fig.\,\ref{f2}, we show results for the quench $\alpha=0.85\to\alpha'=0.1$ as a representative example.  
$|P(t)|$ follows a short time power law in the time window $a<t<\xi$. From the inset
we see that $|P(t)|$ is relatively independent of $\Delta$ at small $t$. 
We see excellent universal scaling in accordance with the Ansatz (\ref{scaling}). 
Note that the scaling factors $\xi$  and $Z$ were not determined by fitting, as
is frequently done when one looks for universality, but were calculated using Eqs.\;(\ref{klexp}) and (\ref{eqz}).  

\begin{figure}[tbh]
\begin{center}
\includegraphics[width=.9\columnwidth]{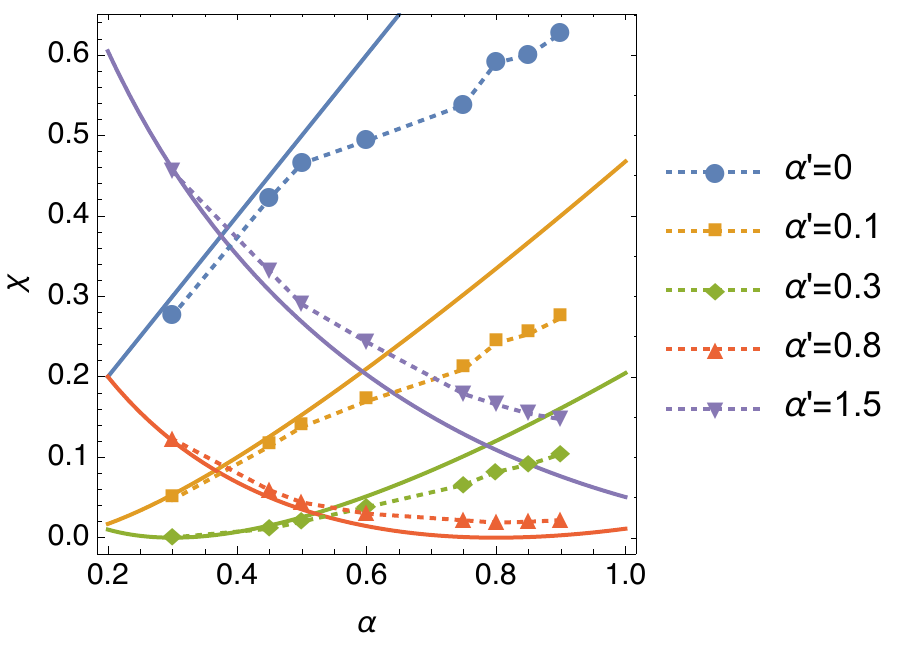}
\caption{Symbols: the short time power law exponent $\chi$ defined in Eq.\;(\ref{spw}), at five different values of $\alpha'$,
as extracted from our numerical $P(t)$ data, an example of which is shown in Fig.\,\ref{f2}.
Curves: the small $\alpha$ prediction $\chi=(\varphi/\pi-\varphi'/\pi)^2$.
\label{f3}}
\end{center}
\end{figure}

In Fig.\,\ref{f3} we show examples of the short time power law exponent we extracted from the Loschmidt echo. 
For $\alpha\lesssim 0.5$, we find good agreement with the approximate analytical result $\chi=(\varphi/\pi-\varphi'/\pi)^2/2$. 
At larger $\alpha$, $|P(t)|$ still  
obeys a short time power law, but the power law exponent deviates from $(\varphi/\pi-\varphi'/\pi)^2/2$. 
At small $\alpha$, modes with $1/\xi\ll \omega_n\ll1/a$ are associated with displacements 
$f_{mn}\simeq\sqrt{\alpha} g_n/2\omega_n$, as if $\Delta=0$, and this allows for a straightforward
explanation of the power law in terms of the response of non-interacting spin-up and spin-down Fermi seas to 
a local quench. However, at larger $\alpha$, we find numerically that the same asymptotic 
behavior of $f_{mn}$ only sets in close to the ultraviolet cutoff $1/a$. 
If a non-interacting Fermi sea explanation is possible in this regime, the coherent state expansion does not make this manifest. 
When $\alpha=\alpha'$, the analytical result of Refs. \cite{Lukyanov1,Lukyanov2} is that $\chi=0$.  The largest deviation
we found from this is $\chi=0.018$ at $\alpha=\alpha'=0.8$. We believe this is a small finite size error, and note that
we unambiguously find that $\chi$ is minimal when $\alpha=\alpha'$.
We also note oscillatory behavior that might be noise, in the results for $\alpha'=0,\,0.1$ and $\alpha>0.6$. 
The same noise is not seen for at other $\alpha'$-values. Given that the same ground state data was used for all $\alpha'$,
we attribute any noise that is present, to the numerical extraction of the power law exponent rather than to errors in the ground state data.
  
\begin{figure}[tbh]
\begin{center}
\includegraphics[width=.9\columnwidth]{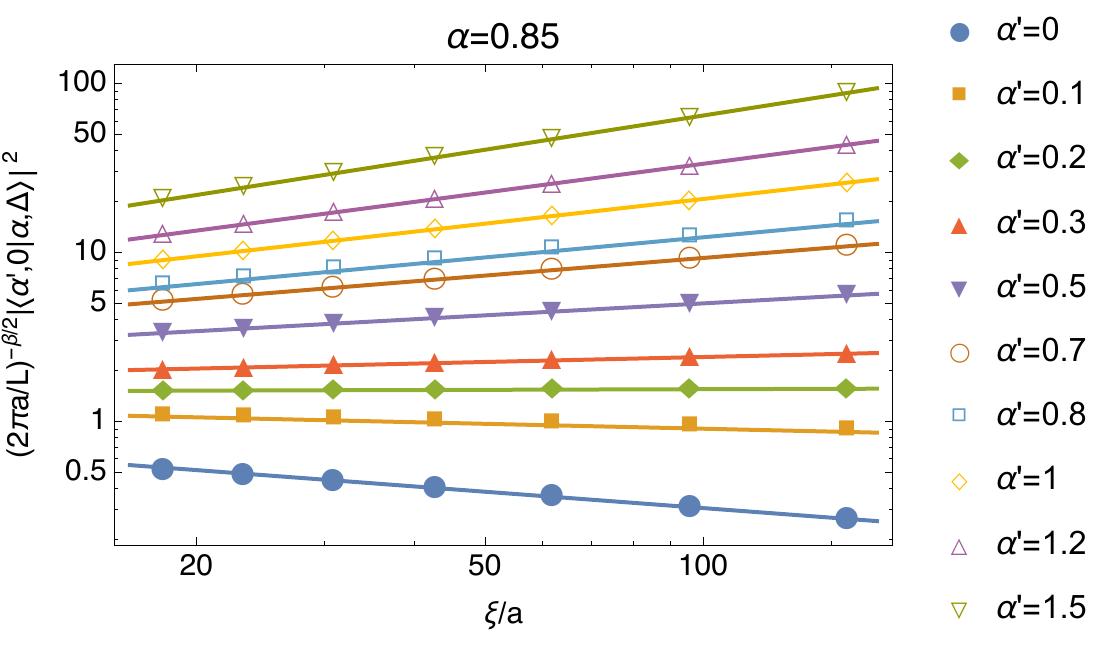}
\caption{Symbols: The ground state to ground state overlap squared $\left|\bra{\alpha',0}\left.\alpha,\Delta\right>\right|^2$ versus the Kondo length
$\xi$ for various $\alpha'$, at $\alpha=0.85$. Lines: the power law
$\left|\bra{\alpha',0}\left.\alpha,\Delta\right>\right|^2=G\xi^\eta$ with $G$ and $\eta$ obtained by fitting to the data with $\xi\gtrsim 25a$.
\label{f4}}
\end{center}
\end{figure}

\begin{figure}[tbh]
\begin{center}
\includegraphics[width=.9\columnwidth]{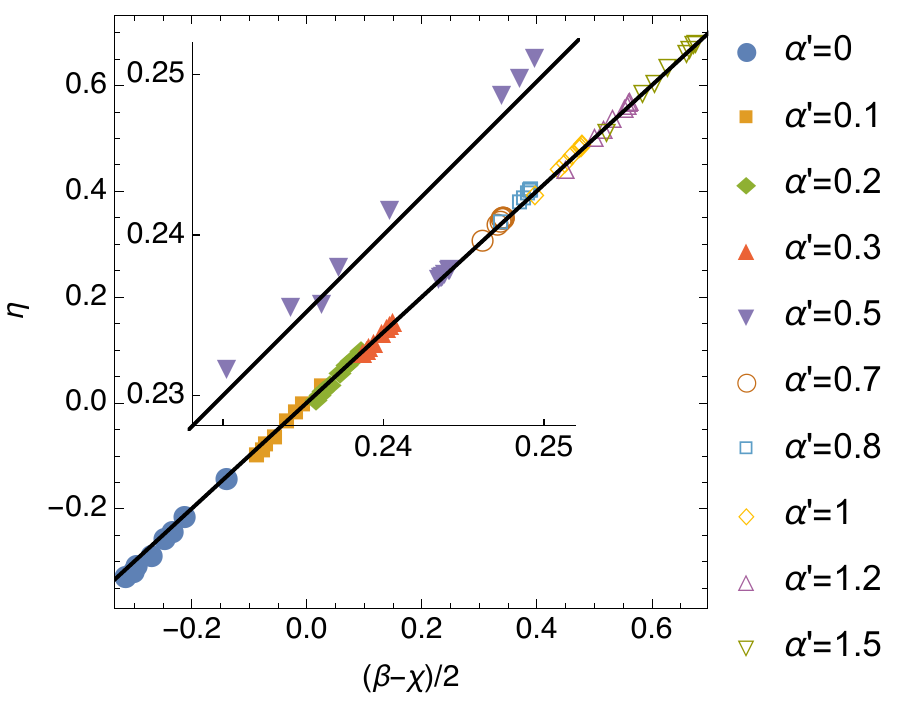}
\caption{Symbols: Each data point corresponds to a fixed value of $\alpha$ and $\beta$. The $x$ and $y$ coordinates of each point is respectively  
$(\beta-\chi)/2$ with the the short time power law exponent $\chi$ corresponding to the symbols in Fig\,\ref{f3}, 
and the power law exponent $\eta$ cf. Eq.\;(\ref{pleta}), as extracted from the data in such as that plotted in Fig.\,\ref{f4}. 
Black line: the heuristic prediction $\eta=(\beta-\chi)/2$.
Inset: an enlarged view of the clustered $\beta=0.5$ data.
\label{f5}}
\end{center}
\end{figure}

With the assumption verified that in the regime $a\ll t\ll \xi$, the Loschmidt echo is both independent of $\Delta$ and obeys a power law $t^{-\chi/2}$, our conclusion
regarding the power law dependence of the ground state overlap on the Kondo length must be correct. 
We have calculated $(2\pi a/L)^{-\beta/2}\left|\bra{\alpha',0}\left.\alpha,\Delta\right>\right|^2$ directly and indeed find that
at fixed $\alpha$ and $\alpha'$ (i.e. fixed $J_\parallel$ and $J_\parallel'$ in the Kondo language)
$\left|\bra{\alpha',0}\left.\alpha,\Delta\right>\right|^2 =Q$ has a power law dependence on $\xi$, i.e.
\begin{equation}
\left|\bra{\alpha',0}\left.\alpha,\Delta\right>\right|^2\propto \xi^{\eta}\label{pleta}.
\end{equation}
As an example, in Fig. \ref{f4}, we show results for $\alpha=0.85$, and various $\alpha'$.
In Fig.\,\ref{f5}, we compare the power law exponent
$\eta$ of the overlap, extracted from data such as that shown in Fig. \ref{f4}, to the short time power law exponent $\chi$ of the Loschmidt echo,
extracted from data such as that shown in Fig. \ref{f2}. 
We see very good agreement with our prediction that $\eta=(\beta-\chi)/2$. The largest deviations are seen at
$\alpha'=0$, $\alpha>0.6$. As we noted already, the extracted values of the exponent $\chi$ for
these data points may contain some noise. However, even here, the largest deviation between the
extracted $\eta$ values and the line representing $\eta=(\beta-\chi)/2$, is $5\%$. For the other data sets,
the deviation is of the order of a percent at most. 

In conclusion, for the Kondo model,
we have demonstrated that the Loschmidt echo, and
the overlap between ground states before and after the quench, are connected not only via dynamics at long times, but also via dynamics at short times. 
Our main results are contained in Eqs.\;(\ref{mainres2}) and (\ref{mainres1}) and can be summarized as follows. If one subtracts the slope seen at
short times in the log-log plot of the Loschmidt echo (Fig. \ref{f2}), from the slope seen at long times, one obtains the slope seen in the log-log plot of 
the ground state to ground state overlap squared versus the Kondo length (Fig. \ref{f4}). The overlap's dependence on the emergent length scale $\xi$
thus elegantly encodes information about both IR and UV dynamics.  Apart from Anderson's orthogonality theorem, the other essential
ingredients that produced this result were universal scaling and power law behavior of the Loschmidt echo at short times. In light of this, it may be interesting to
revisit results for the overlap, in studies that have focussed on system size dependence, to see if similar connections exist.

\acknowledgements
We wish to thank Serge Florens for insightful discussions and comments. This work is based on research supported by the National Research Foundation of South Africa
(Grant Number 90657) and CNRS PICS contract `FermiCats'.

\newpage
\setcounter{figure}{0}
\setcounter{table}{0}
\setcounter{equation}{0}

\onecolumngrid

\global\long\def\theequation{S\arabic{equation}}
\global\long\def\thefigure{S\arabic{figure}}
\renewcommand{\thetable}{S\arabic{table}}
\renewcommand{\arraystretch}{0.6}

\normalsize

\vspace{1.0cm}
\begin{center}
{\bf \large Supplementary material: Ground state overlap knows about ultraviolet dynamics: results for the Kondo model}
\end{center}

\subsection{Identifying the variational parameters of the multi-coherent state expansion}
As mentioned in the main text,
significant headway can be made analytically with the variational optimization of the truncated trial state
\begin{equation}
\ket{\psi}=\sum_{m=1}^M \frac{c_m}{\sqrt{2}}\left(\ket{f_m}\ket{\uparrow}-\ket{-f_m}\ket{\downarrow}\right).\label{trialS}
\end{equation}
Here we briefly review the relevant results.

By setting the variation of  $\left<H_{\alpha,\Delta}\right>$ with respect to $f_{mn}$, $m=\{1,\,2,\,\ldots,\,M\}$,
equal to zero, one obtains a set of equations of the form
\begin{equation}
\left(U \omega_n + V\right)\left(\begin{array}{l}f_{1n}\\ \vdots\\f_{Mn}\end{array}\right)=\sqrt{\alpha} g_n W, \label{vareqS}
\end{equation}
where the real $M\times M$ matrices $U$ and  $V$ and the real column vector $W$ depend on the trial state $\ket{\psi}$ but are 
independent of the mode index $n$, 
i.e. the same $U$, $V$ and $W$ enter the equations for each $n$. (See \cite{SnymanS} for explicit expressions.) 
This structure comes about because (a) $H_{\alpha,\Delta}$ is quadratic in $b_n$ and $b_n^\dagger$, and (b) there are no direct processes in $H_{\alpha,\Delta}$ that scatter
bosons from a mode $n$ to a mode $n'\not=n$. By noting that each matrix element of the inverse of $U \omega_n + V$ is the
ratio of polynomials in $w_n$ that are respectively of order $M-1$ and $M$, and that the denominator is the same for all elements (being the determinant of
$U\omega_n+V$), we discover that
\begin{equation}
f_{m,n}=\frac{\sqrt{\alpha} g_n}{2}h_{m}(\omega_n),~~~h_{m}(z)=\frac{\sum_{l=0}^{M-1}\mu_{m,l}z^l}{\prod_{l=1}^M\left(z-\Omega_l\right)},\label{eqfS}
\end{equation}
where $\mu_{m,l}$ is real. Furthermore, because $U\omega_n+V$ is real, the $\Omega_l$ that are not real come in complex conjugate pairs.  It
is intuitively clear that for modes with natural frequencies $\omega_n \gg \Delta$, the tunnelling term $\Delta\sigma_x/2$ becomes a negligible perturbation and hence
i.e. $f_{mn}\simeq \sqrt{\alpha} g_n/2\omega_n$. This implies that $\mu_{m,M-1}=1$. The same conclusion follows rigorously from the observation
that associated with mode $n$, there are unavoidable positive contributions to the energy proportional to $\omega_n(f_{mn}-\sqrt{\alpha}g_n/2\omega_n)^2$. Energy considerations
also dictate that there are no poles $\Omega_l$ on the positive real line.
Finding the $M^2 + M - 1$ unknown parameters $\Omega_m$, $\mu_{m_1m_2}$ and $c_m$ variationally
constitutes a global minimization problem that we solve numerically. 
However, as we show below, some exact results for quench dynamics can also be derived without explicitly computing the optimal  
$\Omega_m$, $\mu_{m_1m_2}$ and $c_m$.

\subsection{Long time behavior of the Loschmidt echo}

Here we give a simple analytical derivation of the long time behavior of the Loschmidt echo, based on the coherent state expansion.
The derivation does not require explicit computation of the optimal values of the variational parameters, and is valid for an expansion
with an arbitrary number $M$ of terms. We believe the result is therefore exact. 

Firstly note that the ground state energy of the post-quench hamiltonian $H_{\alpha',0}$ is
\begin{equation}
E_0'=-\alpha' \sum_{n=1}^\infty\frac{g_n^2}{4\omega_n},
\end{equation}
Next, note that
because $H_{\alpha',0}$ preserves the $z$-component of spin, and is quadratic in boson operators, it is straightforward to compute the time evolution of $\ket{\alpha,\Delta}$.
After some algebra, these observations lead to the result 
\begin{align}
&P(t)=\sum_{m_1,m_2}c_{m_1}c_{m_2}\bra{f_{m_1}}\left.\alpha'+\right>\left<\alpha'+\right.\ket{f_{m_2}}\nonumber\\
&\times\exp\frac{1}{4}\sum_{n=1}^\infty g_n^2\left[\frac{\sqrt{\alpha'}}{\omega_n}-\sqrt{\alpha}h_{m_1}(\omega_n)\right]\left[\frac{\sqrt{\alpha'}}{\omega_n}-\sqrt{\alpha} h_{m_2}(\omega_n)\right]e^{-i\omega_n t},\label{pt1S}
\end{align}
where
\begin{equation}
\ket{\alpha'\pm}=\exp\pm\sqrt{\alpha'}\sum_{n=1}^\infty \frac{ g_n}{2\omega_n}(b_n^\dagger-b_n)\ket{0},
\end{equation}
so that $\ket{\alpha'+}\ket{\uparrow}$ and $\ket{\alpha'-}\ket{\downarrow}$ are the two degenerate ground states of $H_{\alpha',0}$ and
\begin{equation}
\bra{f_m}\left.\alpha'+\right>=\exp-\frac{1}{8}\sum_{n=1}^\infty g_n^2\left[\frac{\sqrt{\alpha'}}{\omega_n}-\sqrt{\alpha}h_{m}(\omega_n)\right]^2.
\end{equation}

At long times, the frequency sum in Eq.\,(\ref{pt1S}) is dominated by the contribution from small frequencies, where $g_n/\omega_n$ dominates over $h_m(\omega_n)$. 
(Recall that the $h_m(\omega)$ remain finite when $\omega\to0$.)
Thus,
asymptotically
\begin{equation}
P(t)\simeq\left|\left<\alpha,\Delta \right.\left|\alpha',0\right>\right|^2 \exp\frac{\alpha'}{4}\sum_{n=1}^\infty \frac{g_n^2}{\omega_n^2}e^{-i\omega_n t}
\end{equation}
where
\begin{equation}
\ket{\alpha',0}=\left(\ket{\alpha'+}\ket{\uparrow}-\ket{\alpha'-}\ket{\downarrow}\right)/\sqrt{2}
\end{equation}
is the ground state of $H_{\alpha',0}$ with $T$-eigenvalue $-1$. In the thermodynamic limit it is 
convenient to rewrite this result as
\begin{equation}
P(t)\simeq\left|\left<\alpha,\Delta \right.\left|\alpha',0\right>\right|^2\exp\frac{\alpha'}{4}\int_0^\infty d\omega \frac{J(\omega)}{\omega^2}e^{-i\omega t}\label{pt2S}
\end{equation}
using the bath spectral function $J(\omega)$. 

\subsection{Further analytical progress in the Ohmic case}
In the Ohmic case, the sums appearing in the expression for $P(t)$ can be done analytically. Here we discuss this calculation.
We obtain the Ohmic spin-boson model by setting
\begin{equation}
\omega_n=[\hbar v_F]q_n,~~~q_n=\frac{2\pi n}{L},~~~n=1,\,2,\,3,\,\ldots
\end{equation}
where the infrared scale is introduced via a (large) system size $L$.  
The requirement that the bath should be Ohmic then implies
\begin{equation}
g_n=2\sqrt{\frac{\pi q_n}{L}}e^{-aq/2},
\end{equation}
where we set the ultraviolet scale via a short distance cut-off $a\ll L$.
It is useful to write the Loschmidt echo (\ref{pt1S}) as
\begin{equation}
P(t)=\sum_{m_1m_2}c_{m_1}c_{m_2}\bra{f_{m_1}}\left.\alpha'+\right>\left<\alpha'+\right.\ket{f_{m_2}}P_{m_1m_2}(t),\label{lo1S}
\end{equation}
where 
\begin{align}
&P_{m_1m_2}(t)=\nonumber\\
&\exp\sum_{n=1}^\infty \frac{\pi q_n e^{-(a+it) q_n}}{L}
\bigg\{\alpha h_{m_1}(q_n)h_{m_1}(q_n)
-\frac{\sqrt{\alpha\alpha'}}{q_n}\left[h_{m_1}(q_n)+h_{m_2}(q_n)\right]
+\frac{\alpha'}{q_n^2}
\bigg\}.
\end{align}
The first and second terms in the curly brackets need no infrared regularization and we can replace the sum over discrete modes with an integral. We  cannot do the
same with the third term. Here however, the discrete sum is computed straightforwardly.
\begin{equation}
\sum_{n=1}^\infty \frac{\pi \alpha' e^{-a q_n}}{L q_n}e^{-(a+it)q_n}=-\frac{\alpha'}{2}{\rm ln}\left[1-e^{-2\pi(a+it)/L}\right].
\end{equation}   
Thus 
\begin{align}
&P_{m_1m_2}(t)=\nonumber\\
&\left[1-e^{-2\pi(a+it)/L}\right]^{-\alpha'/2}
\exp\left\{\frac{\alpha}{2}A_{m_1m_2}(t)-\frac{\sqrt{\alpha\alpha'}}{2}\left[B_{m_1}(t)+B_{m_2}(t)\right]\right\},\label{lo2S}
\end{align}
where
\begin{equation}
A_{m_1m_2}(t)=\int_0^\infty dq\, qe^{-q(a+it)}h_{m_1}(q)h_{m_2}(q),~~B_{m}(t)=\int_0^\infty dq\, e^{-q(a+it)}h_{m}(q).
\end{equation}
Also
\begin{equation}
\left<\alpha'+\right.\ket{f_{m}}=\exp\left[-\frac{\alpha}{4}A_{mm}(0)+\frac{\sqrt{\alpha\alpha'}}{2}B_m(0)\right]\left(\frac{2\pi a}{L}\right)^{\alpha'/4}.\label{olS}
\end{equation}

The fact that $h_m(q)$ is a rational function of $q$ allows us to perform the above integrals analytically. Using a method explained in detail in \cite{SnymanS}, we obtain
\begin{align}
A_{m_1m_2}(t)&=e^{-i\phi}\sum_{l=1}^M {\rm Res}\left[q e^{-i\phi}h_{m_1}(qe^{-i\phi})h_{m_2}(qe^{-i\phi})F(q|a+it|),q=\Omega_l e^{i\phi}\right]\nonumber\\
&=\sum_{l=1}^M\Bigg\{\left[h_{m_1l}h_{m_2l}+\Omega_l\left(\dot{h}_{m_1 l}h_{m_2 l}+h_{m_1 l}\dot{h}_{m_2 l}\right)\right]F(\Omega_l(a+it))\nonumber\\
&-\Omega_l(a+it)h_{m_1l}h_{m_2l}\left[F(\Omega_l(a+it))+\frac{1}{\Omega_l(a+it)}\right]\Bigg\}
\end{align}
and
\begin{align}
B_m(t)&=e^{-i\phi}\sum_{l=1}^M {\rm Res}\left[h_{m_1}F(q|a+it|),q=\Omega_l e^{i\phi}\right]\nonumber\\
&=\sum_{l=1}^M h_{m_1l}F(\Omega_l(a+it))\label{lo3S}
\end{align}
where $F(z)=e^{-z}\Gamma(0,-z)$ and $\Gamma(\gamma,z)$ is the incomplete Gamma function. 
The phase $\phi$ is the argument of $a+it$, i .e.
$e^{i\phi}=(a+it)/|a+it|$, while 
\begin{equation}
h_{ml}=\lim_{q\to\Omega_l}(q-\Omega_l)h_m(q)=\frac{\sum_{n=0}^{M-1}\mu_{mn}\Omega_l^n}{\prod_{n=1\not=l}^M(\Omega_l-\Omega_n)},\label{lo4S}
\end{equation}
and
\begin{align}
\dot{h}_{ml}&=\lim_{q\to\Omega_l}\frac{d}{dq}\left[(q-\Omega_l)h_m(q)\right]\nonumber\\
&=\frac{\sum_{n=1}^{M-1}\mu_{mn}n\Omega_l^{n-1}}{\prod_{n=1\not=l}^M(\Omega_l-\Omega_n)}-h_{ml}\sum_{n=1\not=l}^M\frac{1}{\Omega_l-\Omega_n}.\label{lo5S}
\end{align}
In these formulas we assume that the poles $\Omega_l$
of $h_m(q)$ all have negative real parts, a fact that we have not proven. It is however borne out by the numerics. 
Furthermore, the additional terms that would occur if there were poles
with positive real parts, would only introduce additional contributions to the Loschmidt echo that decay exponentially over time, and are unimportant at large times. 
These equations form the basis of our numerical study of the full $P(t)$. 
For future reference, we express the the Kondo length $\xi$ and the scaling factor $Z$, as introduced in the main text, in terms of the ground state parameters. 
\begin{align}
\xi&=\frac{\sqrt{\alpha}}{2}\sum_{m_1m_2}c_{m_1}c_{m_2}\left[h_{m_1}(0)+h_{m_2}(0)\right]\left<f_{m_1}\right|\left.f_{m_2}\right>.\label{klexpS}
\end{align}
\begin{align}
Z&=\left(\frac{2\pi \xi}{L}\right)^{-\alpha'/2}\left|\bra{\alpha',0}\left.\alpha,\Delta\right>\right|^2\nonumber\\
&=\left(\frac{\xi}{a}\right)^{-\alpha'/2}\left|\sum_{m=1}^M c_m\exp\left[-\frac{\alpha}{4}A_{mm}(0)+\frac{\sqrt{\alpha\alpha'}}{2}B_m(0)\right]\right|^2.
\label{eqzS}
\end{align}

\subsection{Analytical results at small $\alpha$}
The coherent state expansion truncated to a single term ($M=1$) is known as the Silbey-Harris Ansatz \cite{SilbeyS,HarrisS}. The Silbey-Harris Ansatz is exact for small $\alpha$.
Here we use this fact to obtain an explicit expression for $P(t)$ in the limit of small $\alpha$. We use this expression to give an analytical derivation of the short time power law
of $P(t)$, valid for sufficiently small $\alpha$. For $M=1$, there is only one variational parameter $\Omega_1$. Its optimal value is $-\Delta_R$ where $\Delta_R$ satisfies
\begin{equation}
\Delta_R=\Delta\left<f_1\right.\ket{-f_1}=\Delta \exp-\alpha A_{11}(0)<\Delta
\end{equation}
The Kondo length is
\begin{equation}
\xi=\frac{\sqrt{\alpha}}{\pi\Delta_R}.
\end{equation}  
For $A_{11}(t)$ and $B_{1}(t)$ we obtain
\begin{eqnarray}
A_{11}(t)&=&\left[1+\Delta_R(a+it)\right]F(-\Delta_R(a+it))-1\nonumber\\
B_1(t)&=&F(-\Delta_R(a+it))
\end{eqnarray}
and for $P(t)$ we obtain
\begin{equation}
P(t)=\left(\frac{it}{a}\right)^{-\alpha'/2}\exp\left\{\frac{\alpha}{2}\left[A_{11}(t)-A_{11}(0)\right]-\sqrt{\alpha\alpha'}\left[B_1(t)-B_1(0)\right]\right\}
\end{equation}
Provided $\Delta$ is sufficiently smaller than $1/a$, we have $\Delta_R a,\,\Delta_Rt \ll 1$ in the regime $a\ll t<\ll \xi$ and we can expand $F(-\Delta_R(a+it))$ for
small arguments using $F(-z)=-\ln(z)-\gamma_E+\mathcal O(z)$ where $\gamma_E=0.57\ldots$ is the Euler-Mascheroni constant. 
We then find $\left[A_{11}(t)-A_{11}(0)\right]=\left[B_1(t)-B_1(0)\right]=-\ln(1+it/a)$ independent of $\xi$, and 
 \begin{equation}
 P(t)=\left(\frac{it}{a}\right)^{-\chi/2},
 \end{equation}
with $\chi=(\sqrt{\alpha}-\sqrt{\alpha'})^2$. We note that the same result would be obtained if we took $f_{1n}=\sqrt{\alpha}g_n/2\omega_n$. In the Kondo language this 
corresponds to the initial state 
$\ket{\rm UV}=(\ket{-\varphi}\left|\uparrow\right>-\ket{\varphi}\left|\downarrow\right>)/\sqrt{2}$, with $\varphi=\pi(1-\sqrt{\alpha})$. 
Here $\ket{\pm \varphi}$
represents a noninteracting Fermi sea in which electrons undergo phase shifts that depend on
their spin direction, such that at the Fermi energy these phase shifts equal $\pm\varphi/2$
and $\mp\varphi/2$ for spin up and spin down electrons respectively.

\subsection{Kondo model and Andersons orthogonality theorem}
 
The algebraic decay $t^{-\alpha'/2}$ and $L^{-\alpha'/2}$ of respectively the Loschmidt echo and the overlap has been obtained
before, for instance using numerical renormalization group technology \cite{WeichselbaumS}. It is a manifestation of the fact that
under renormalization, the Kondo ground state (of the pre-quench hamiltonian)
flows to a strong coupling infrared fixed point of the from 
$\ket{\rm IR}=(\ket{\theta=-\pi}\left|\uparrow\right>-\ket{\theta=\pi}\left|\downarrow\right>)/\sqrt{2}$. Here, we use the notation $\ket{\theta}$ to
represent a noninteracting Fermi sea in which electrons undergo phase shifts that depend on
their spin direction, such that at the Fermi energy these phase shifts equal $\theta/2$
and $-\theta/2$ for spin up and spin down electrons respectively. The post-quench ground state on the other hand, is of the form
$\ket{\varphi',\infty}=(\ket{\theta=-\varphi'}\left|\uparrow\right>-\ket{\theta=\varphi'}\left|\downarrow\right>)/\sqrt{2}$.
The power law behavior 
of the overlap can be understood by replacing the true initial state with $\ket{\rm IR}$ and applying the celebrated
Anderson orthogonality theorem, which states that $\left|\left<\pm\pi\right.\ket{\pm\varphi'}\right|^2\propto (L/\lambda_F)^{-(1/2-\varphi'/2\pi)^2}$
where $\lambda_F$ is the Fermi wavelength and the proportionality constant is of order unity. Since the infrared fixed point of the Kondo model is reached when
length scales shorter than $\sim\xi$ are integrated out, the effective Fermi wavelength to use when applying the Anderson orthogonality theorem is of order $\xi$. 
Another classic result in many-body theory \cite{OthakaS} states that 
for non-interacting fermions, the Loschmidt echo decays algebraically $\propto (\varepsilon_F t)^{-(1/2-\varphi'/2\pi)^2}$, with the same power law exponent
as that governing the $L$-dependence of the overlap. When applying this result to the long time dynamics of the Kondo model, the effective Fermi energy
to use is $\varepsilon\sim \hbar v_F/\xi$. Finally we note that the coherent state expansion provides an analytic derivation of these
results. The derivation is exact because no restriction is placed on the number $M$ of terms to which the expansion is truncated.

\subsection{Convergence of the multi-polaron expansion}
\begin{figure}
\begin{center}
\includegraphics[width=.75\textwidth]{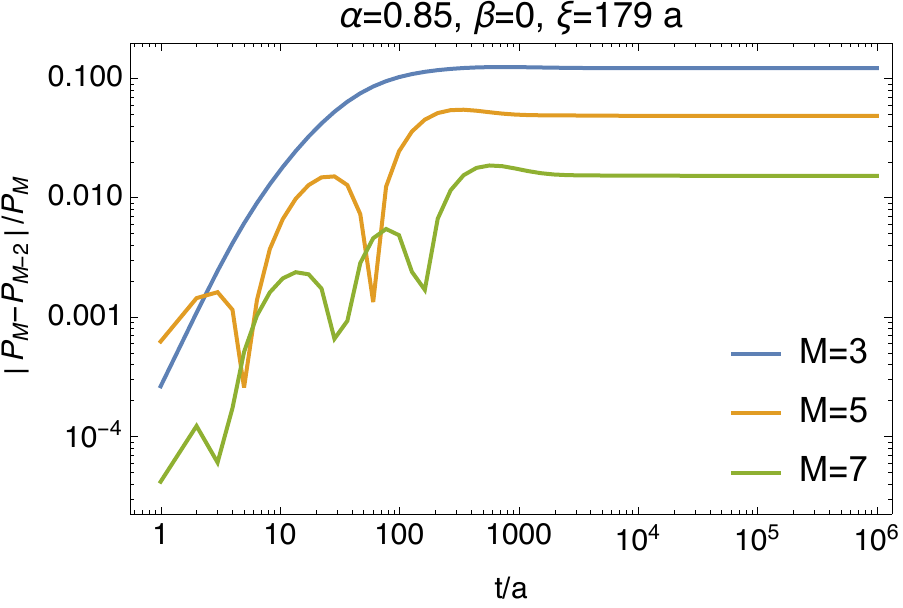}
\caption{The relative change in $P(t)$ when the number of terms in the trial state is incremented by $2$. Results are shown for $\alpha=0.85$ and
$\Delta=0.157 a$. We estimate $\xi=180 a$. We show curves for $\alpha'=0$.\label{f1s}}
\end{center}
\end{figure}

In Ref. \cite{SnymanS} numerous tests were performed to demonstrate that a sufficient number of terms were included to ensure good convergence
of the numerically obtained ground state data. Here we provide one more piece of evidence. For given $\alpha$ and $\Delta$, we optimize multi-polaron trial
states with successively larger $M$. Because of parity effects, we increase $M$ by two in each step. 
We then calculate the relative change in $P(t)$ as $M$ is increased. In Fig.\,\ref{f1s}, we show the results obtained at the least converged datapoint in our
full data set, namely $\alpha=0.85$, $\Delta=0.157 a$. We consider $\alpha'=0$, which corresponds to the case where the extracted short-time power law
exponent is likely least accurate. We see that convergence is better for smaller $t$. This is easy to understand: optimizing ultraviolet degrees of freedom
is prioritized, because this yields greater energy gains than optimizing infrared degrees of freedom. We also see the rapid convergence of the Ansatz: The change
in $P$ drops by roughly $\sqrt{10}$ between successive increments of $M$ by $2$. We therefore expect the $P(t)$ curve for $M=7$ to be accurate to about $1\%$, with
a higher accuracy at short times.

\begin{figure}[h]
\begin{center}
\includegraphics[width=.7\columnwidth]{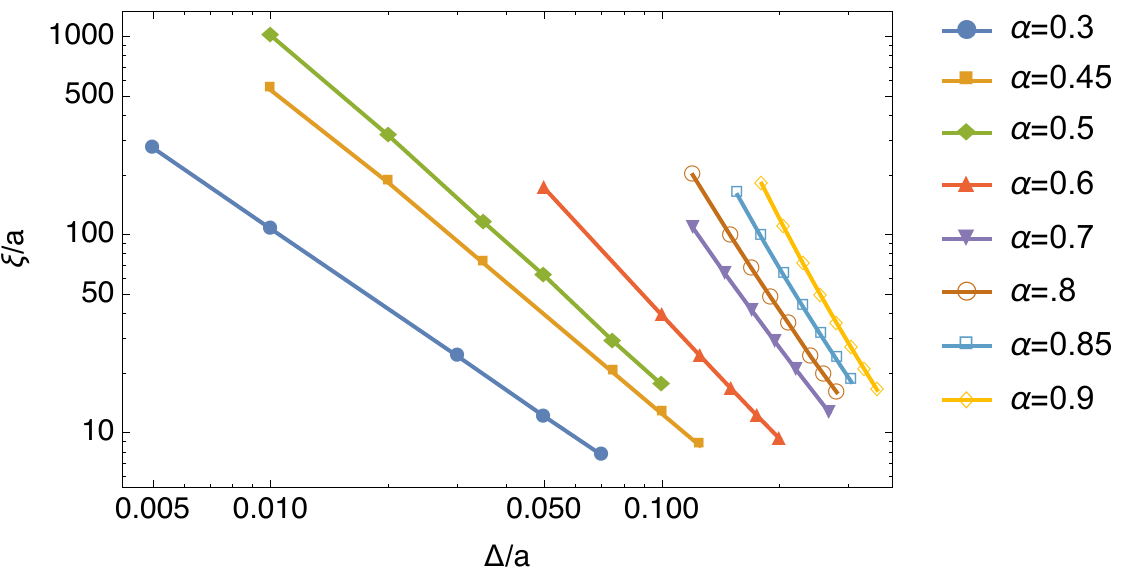}
\caption{The calculated Kondo length  $\xi$ for each of the parameter values $(\alpha,\Delta)$
for which the ground state of $H_{\alpha,\Delta}$ was numerically calculated. Lines connect data points
corresponding to the same $\alpha$. \label{f2s}}
\end{center}
\end{figure}

\begin{figure}[h]
\begin{center}
\includegraphics[width=.32\columnwidth]{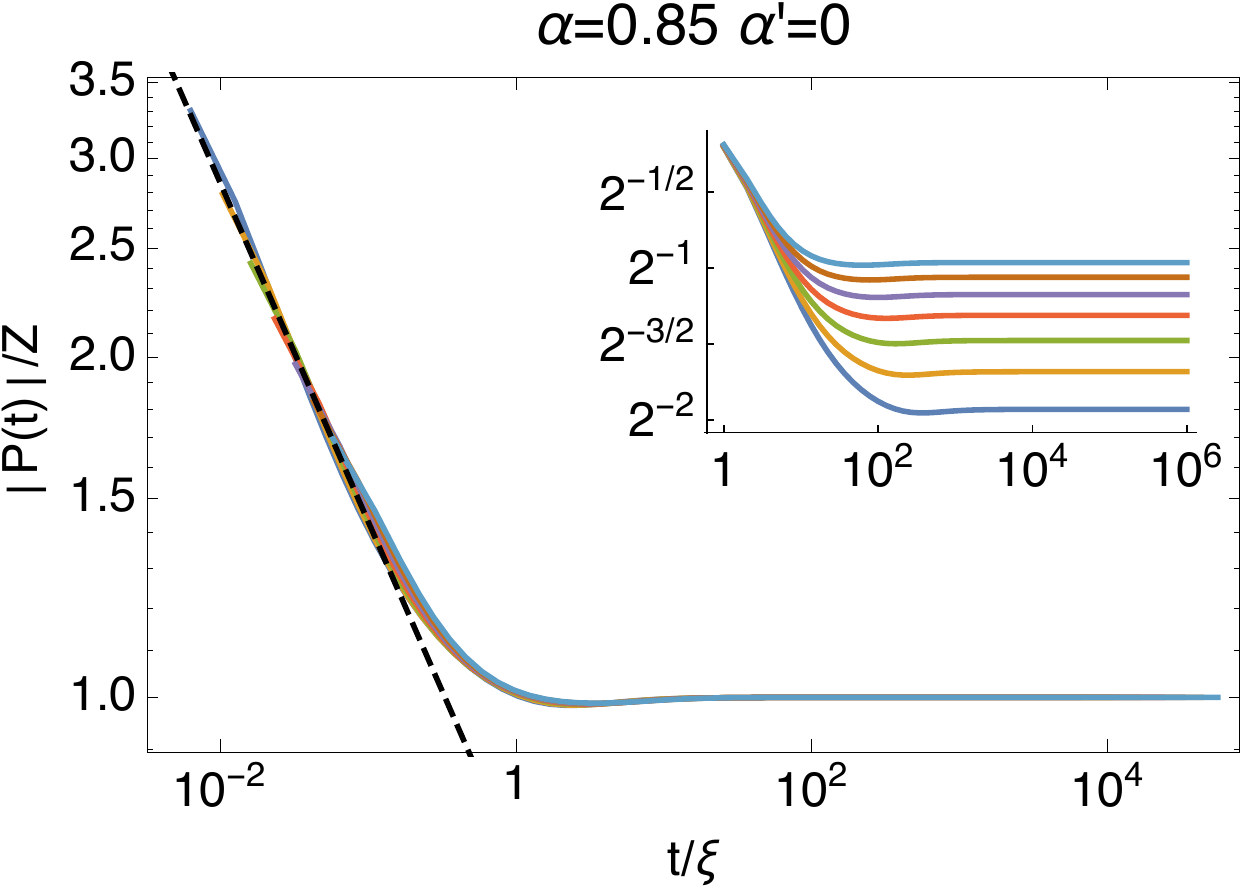}
\includegraphics[width=.32\columnwidth]{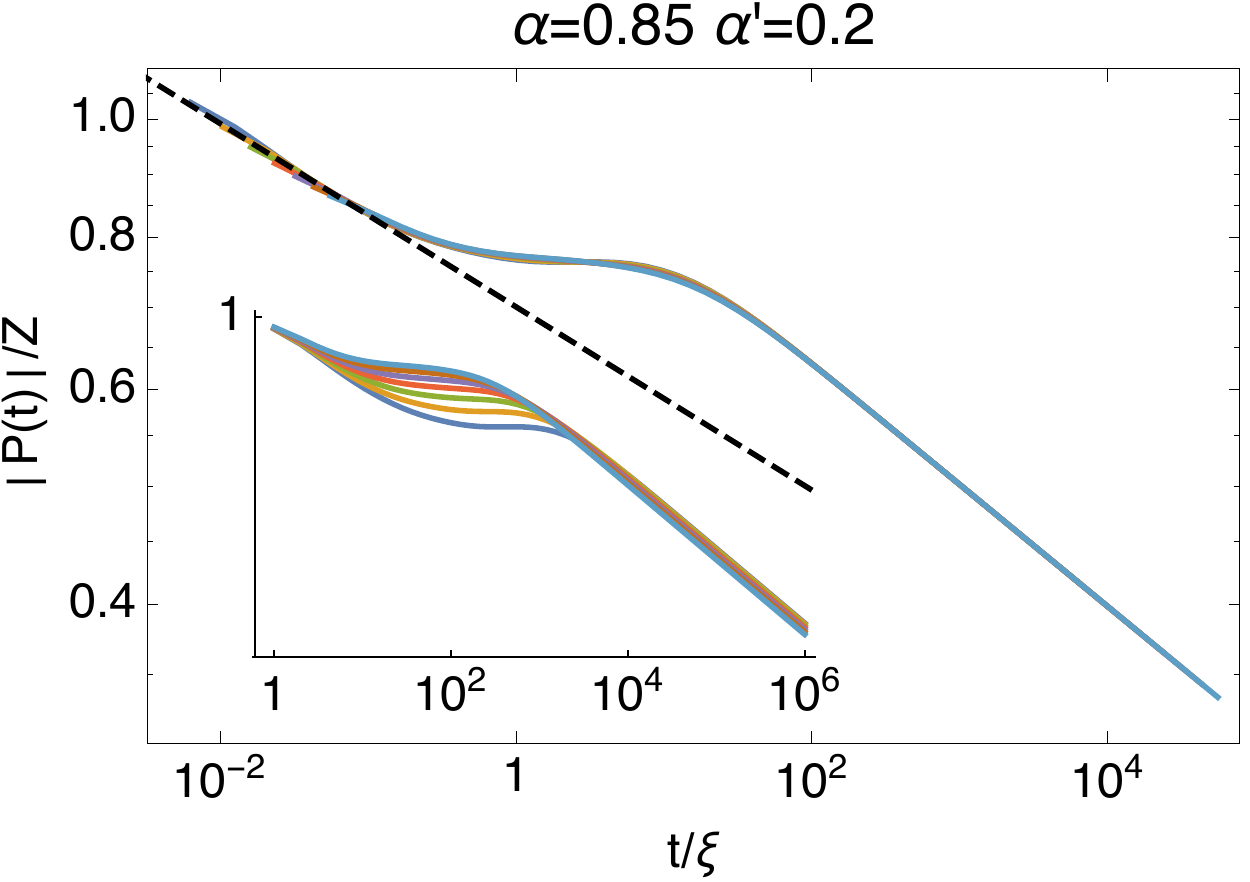}
\includegraphics[width=.32\columnwidth]{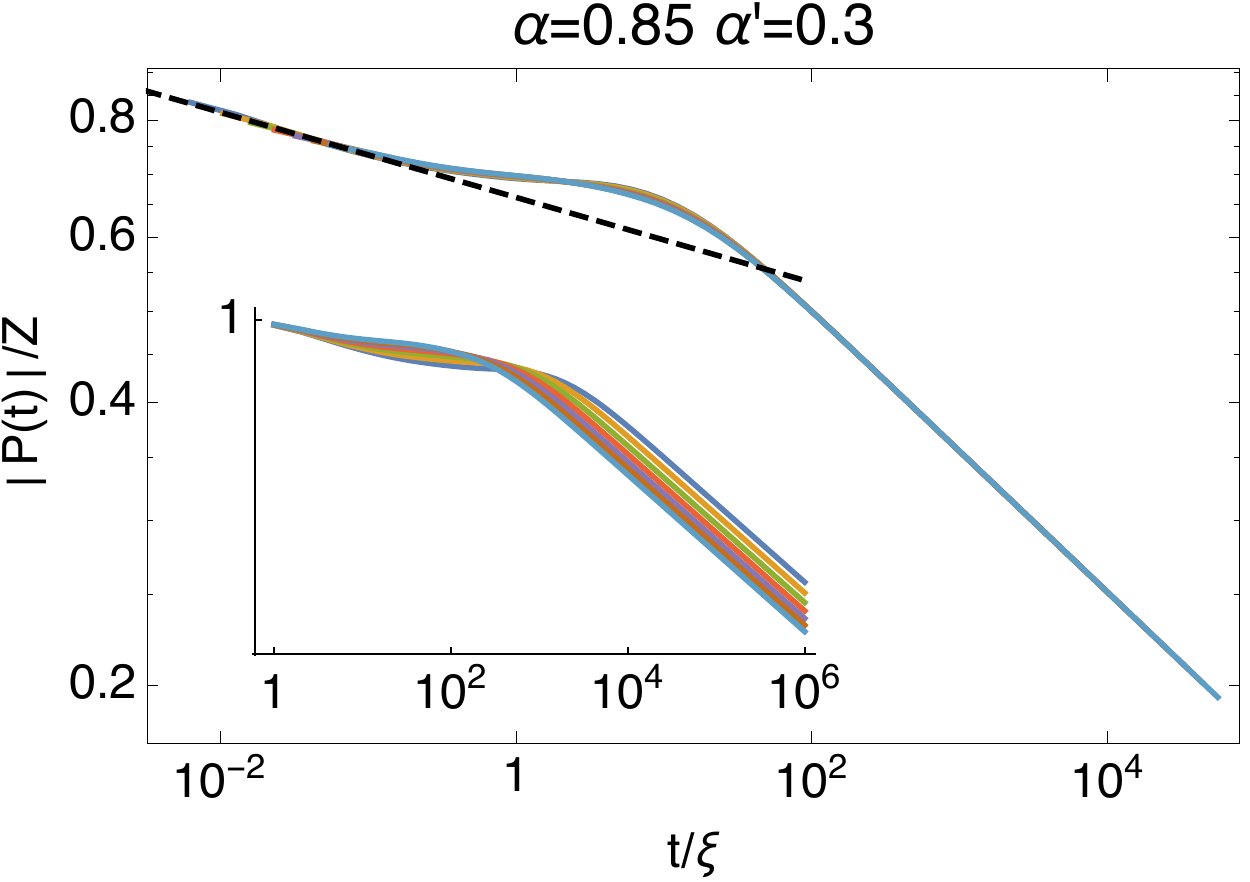}
\includegraphics[width=.32\columnwidth]{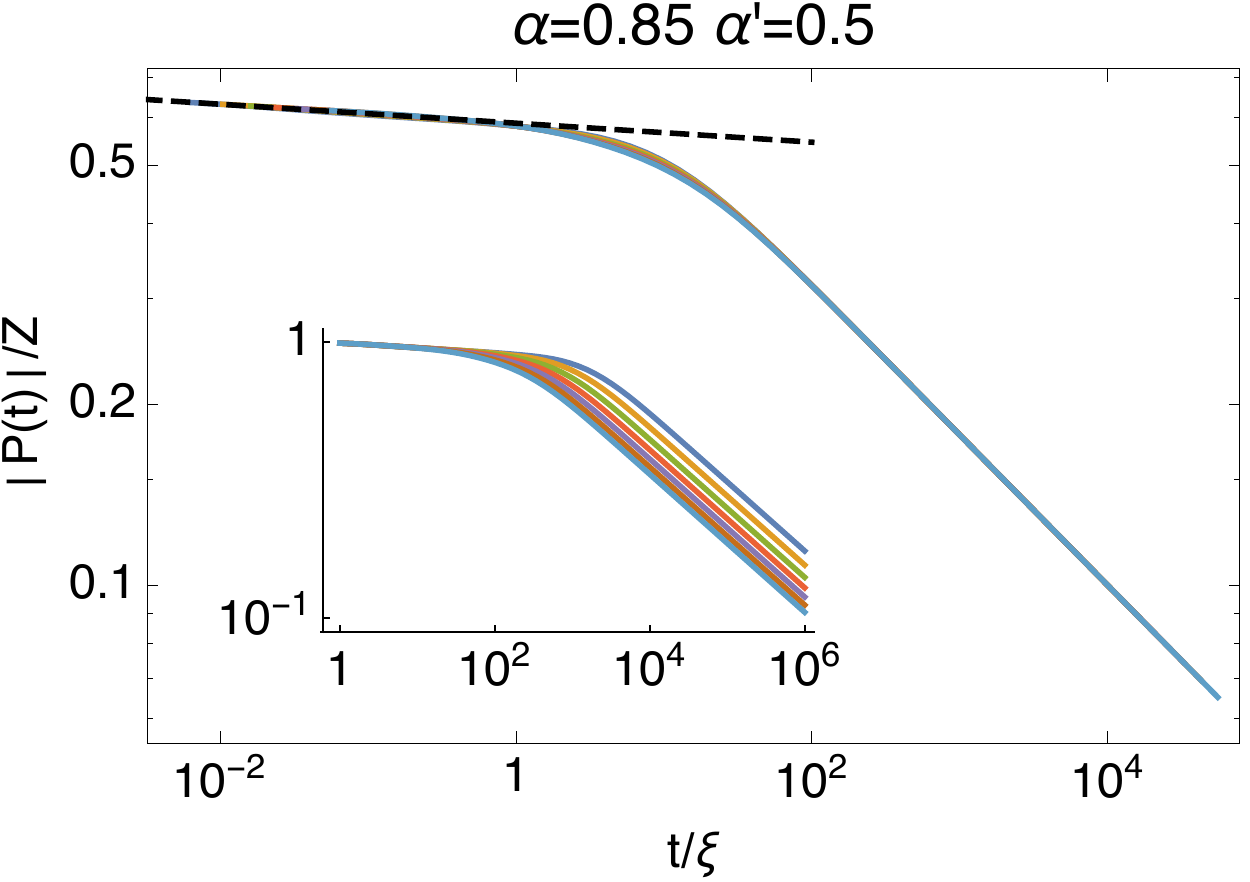}
\includegraphics[width=.32\columnwidth]{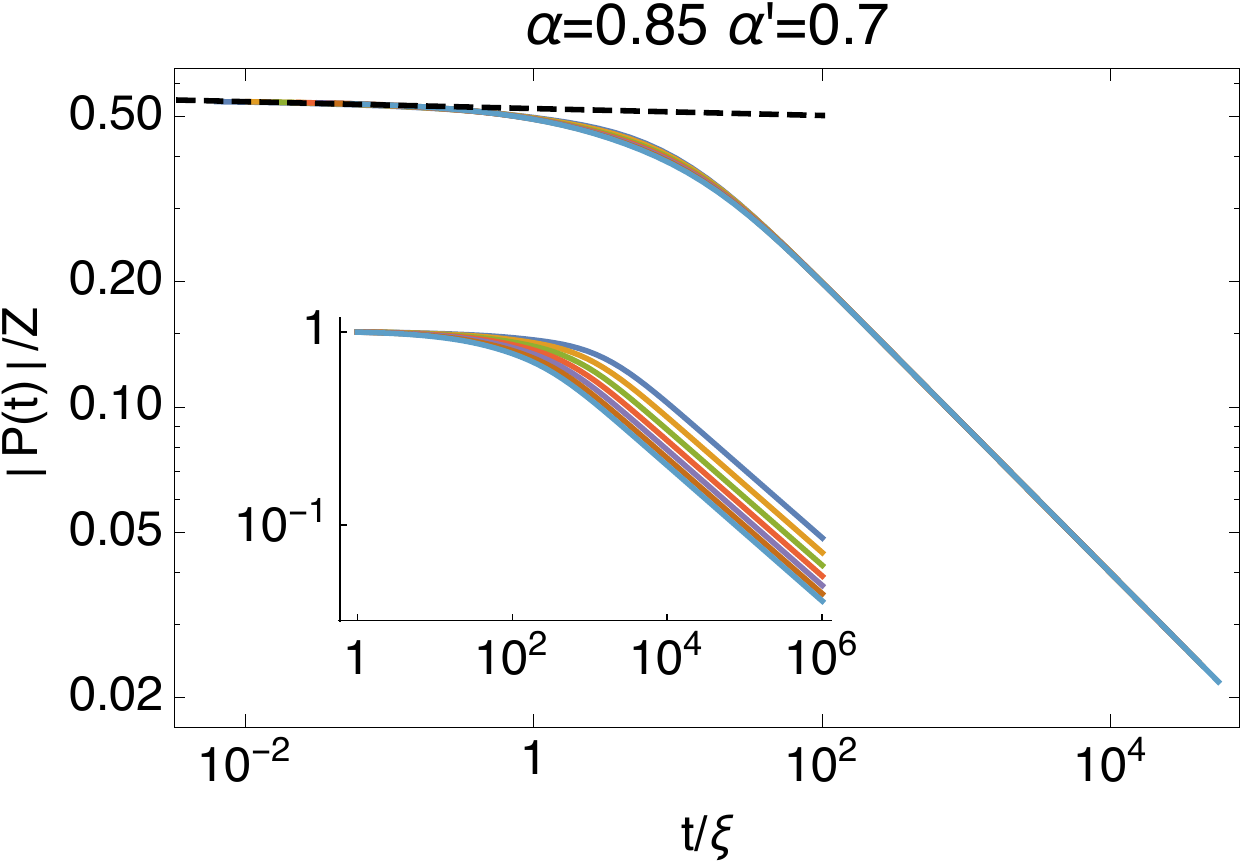}
\includegraphics[width=.32\columnwidth]{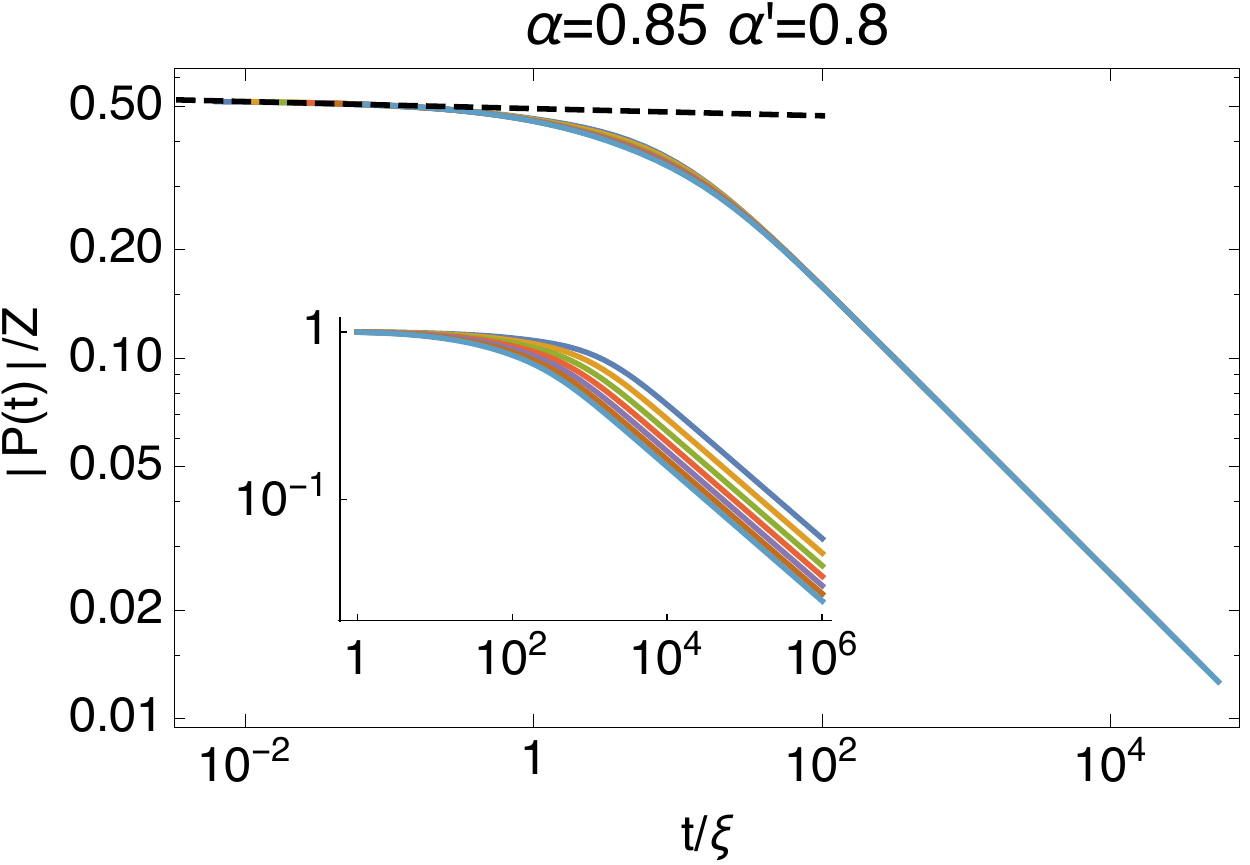}
\includegraphics[width=.32\columnwidth]{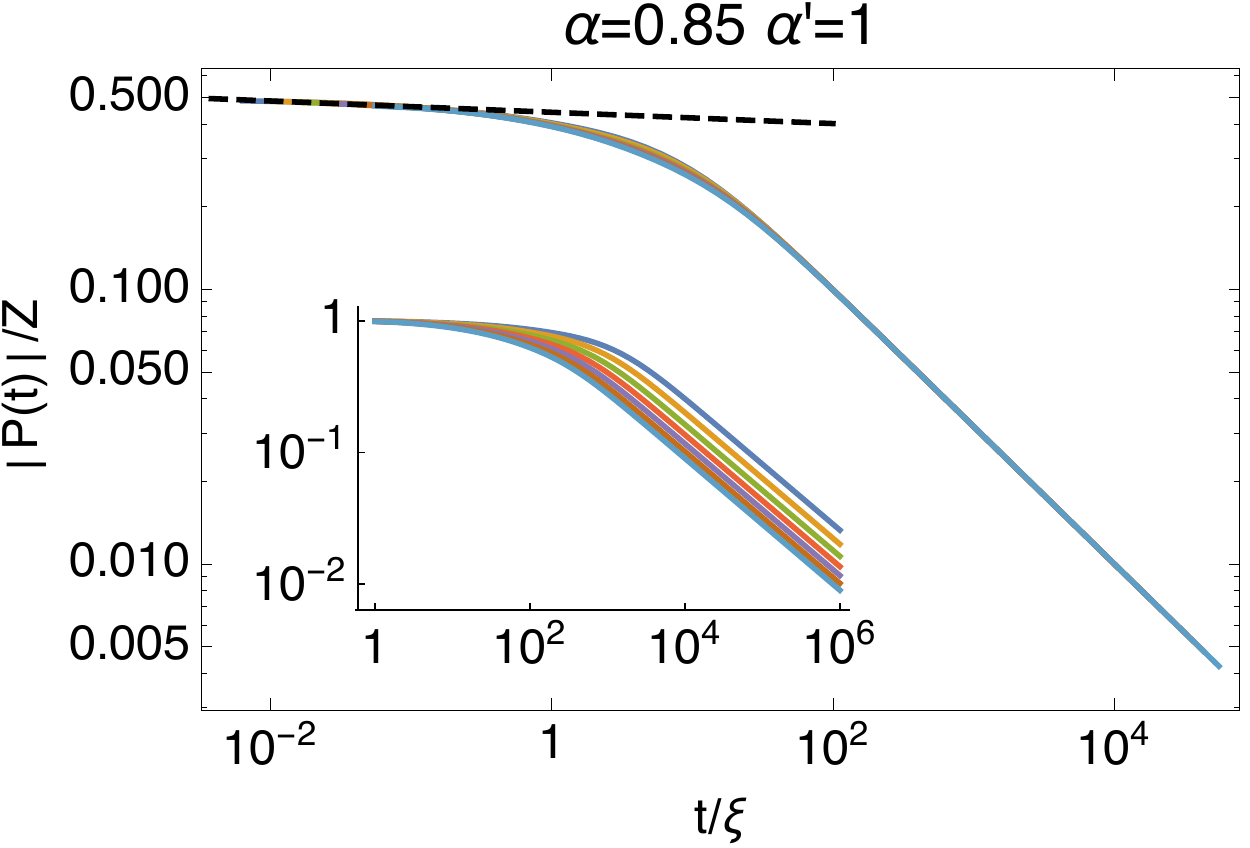}
\includegraphics[width=.32\columnwidth]{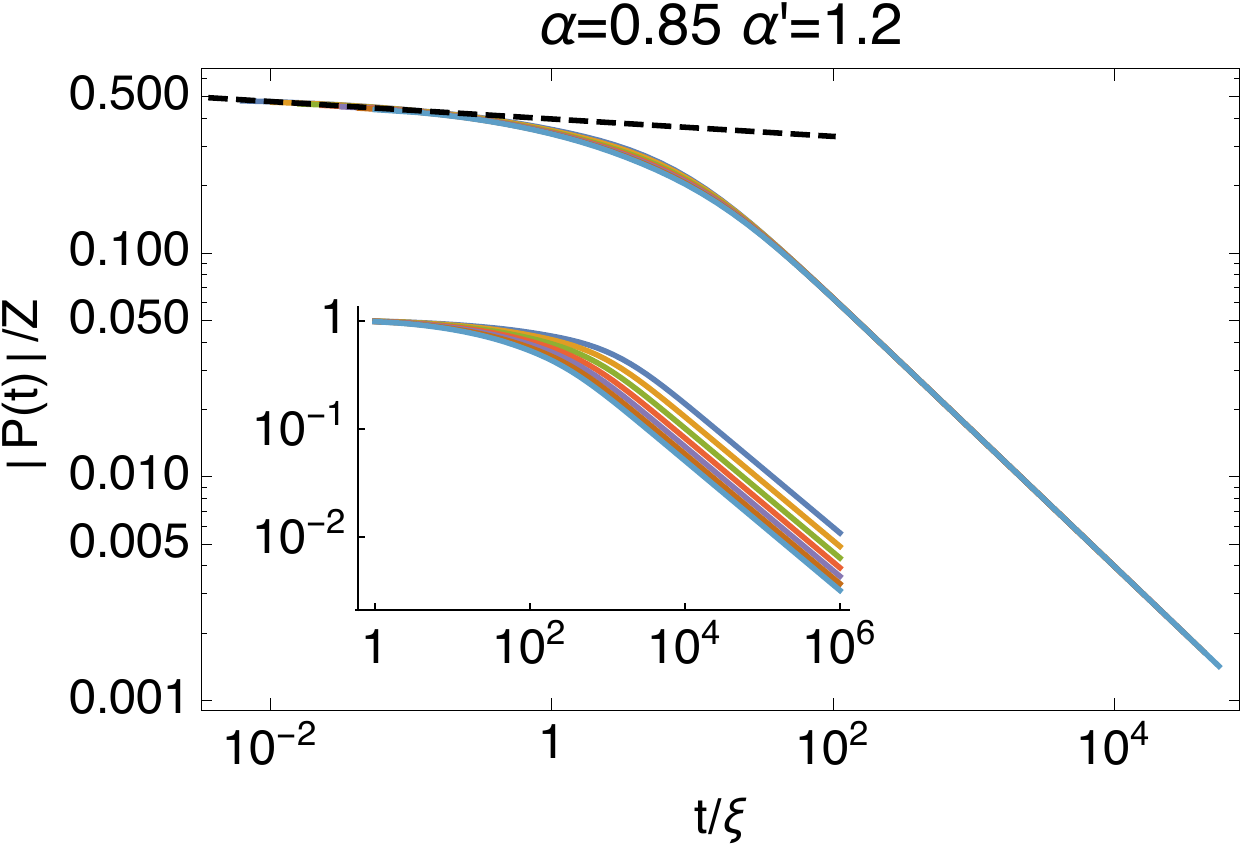}
\includegraphics[width=.32\columnwidth]{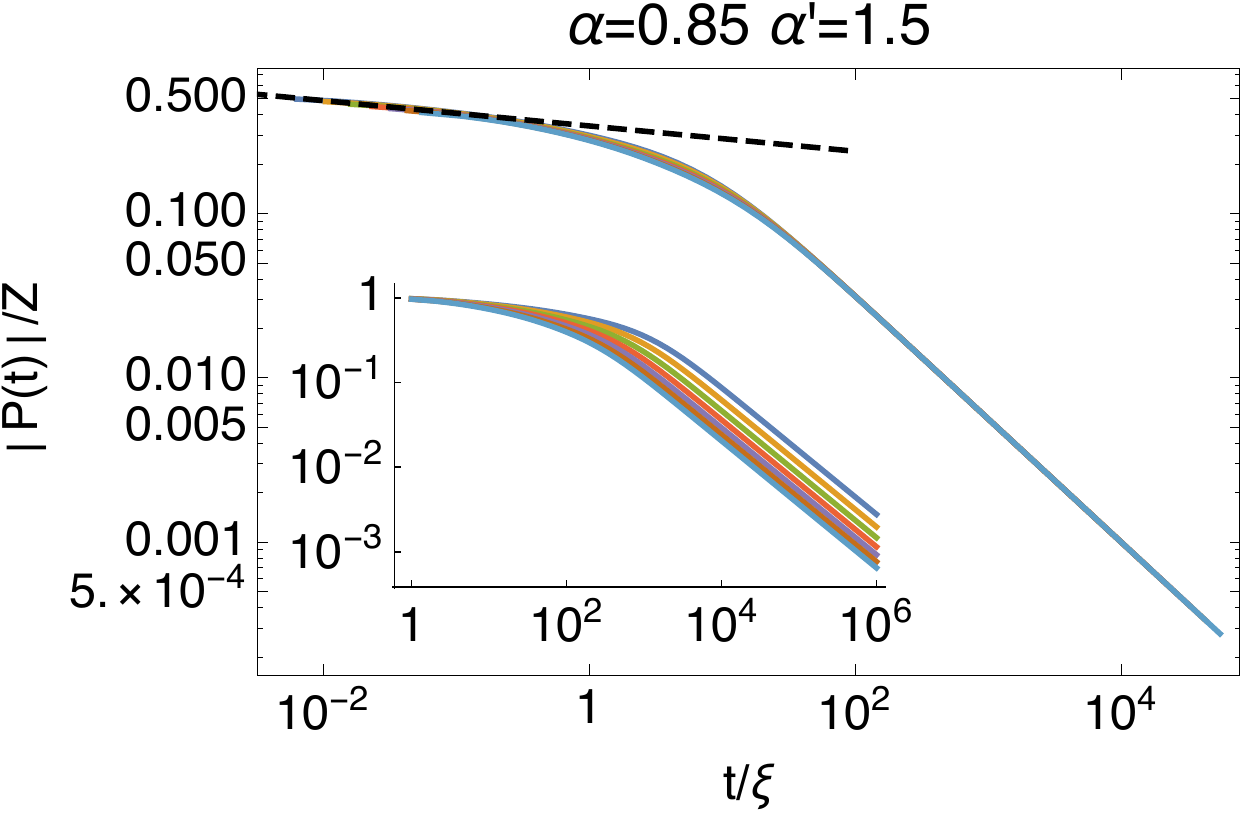}
\caption{The amplitude of the computed Loschmidt echo $P(t)$ for $\alpha=0.85$. Different panels correspond to different $\alpha'$ values The different curves correspond different $\Delta/a$ values
$\in\{0.156,\,0.18,\,0.205,\,0.23,\,0.255,\,0.28,\,0.305\}$. The main panel shows the amplitude of the scaled function
$P(t)/Z$ v. $t/\xi$ with $Z$ and $\xi$ calculated from Eqs. (\ref{klexpS}) and (\ref{eqzS}). The inset shows the unscaled data v. time in
units of $a$. The sloped dashed line represents the short time power law $|P|=C(t/a)^{-\xi/2}$ with $C$ and $\chi$
obtained by fitting to the amplitude data for $t<10^{-1}\xi$. \label{f3s}}
\end{center}
\end{figure}

\subsection{Further numerical results}

Here we provide some further numerical results. In Fig.\,\ref{f2s} we plot the Kondo lengths associated with each of the ground states we obtained numerically.
In Fig.\,\ref{f3s} we plot the rest of numerical Loschmidt echo data for the $\alpha=0.85$ data set. (The data for $\alpha'=0.1$ was already presented in the main text.)

\end{document}